\documentclass{article}

\usepackage{arxiv}
\usepackage[utf8]{inputenc} 
\usepackage[T1]{fontenc}    
\usepackage{hyperref}       
\usepackage{url}            
\usepackage{booktabs}       
\usepackage{amsfonts}       
\usepackage{nicefrac}       
\usepackage{microtype}      
\usepackage{lipsum}
\usepackage{graphicx}

\title{Analytic traveling-wave solutions of the Kardar-Parisi-Zhang interface growing equation with different kind of noise terms}

\author{
  Imre Ferenc Barna \\
  Wigner Research Center\\
  Hungarian Academy of Sciences\\
  H-1525 Budapest, P.O. Box 49, Hungary \\
  \texttt{barna.imre@wigner.mta.hu} \\
   \And
 Gabriella Bogn\'ar \\
  Institute of Machine and Product Design\\
  University of Miskolc\\
  Miskolc-Egyetemv\'aros 3515, Hungary \\
  \And
   Mohammed Guedda\\
  Faculte de Mathematiques et d'Informatique\\
  Universit\'e de Picardie Jules Verne Amiens\\
  33, rue Saint-Leu 80039 Amiens, France \\
  \And
  Kriszti\'an Hricz\'o \\
  Institute of Mathematics\\
  University of Miskolc\\
  Miskolc-Egyetemv\'aros 3515, Hungary \\
  \And
  L\'aszl\'o  M\'aty\'as \\
  Department of Technical and Natural Sciences\\
  Sapientia University\\
  Libert\u{a}tii sq. 1, 530104 Miercurea Ciuc, Romania \\
 }

\begin{document}
\maketitle

\begin{abstract}
The one-dimensional Kardar-Parisi-Zhang dynamic interface growth equation
with the traveling-wave Ansatz is analyzed. As a new feature additional analytic
terms are added. From the mathematical point of view, these can be considered as
various noise distribution functions. Six different cases were investigated among others Gaussian,
Lorentzian, white or even pink noise. Analytic solutions are evaluated and
analyzed for all cases. All results are expressible with various special functions
Mathieu, Bessel, Airy or Whittaker  functions showing a very rich mathematical structure with some common general characteristics.
This study is the continuation of our former work, where the same physical phenomena was investigated with the self-similar Ansatz.
The differences and similarities among the various solutions are enlightened.
\end{abstract}


\section{Introduction}
\label{sec:1}
Solidification fronts or crystal growth is a scientific topic which
attracts  much interest from a long time. Basic physics of growing
crystallines can be found in large number of  textbooks (see e.g.,
\cite{konyv}). One of the simplest nonlinear generalization of the
ubiquitous diffusion equation is the so called Kardar-Parisi-Zhang
(KPZ) model obtained from Langevin equation
\begin{equation}
\frac{\partial u}{\partial t} = \nu {\bf{\nabla}}^2 u +
\frac{\lambda}{2}({\bf{\nabla}} u)^2 + \eta({\bf{x}},t), \label{kpz}
\end{equation}
where $u$ stands for the profile of the local growth \cite{kpz}. The
first term on the right hand side describes relaxation of the
interface by a surface tension preferring a smooth surface.  The
next term is the lowest-order nonlinear term that can appear in
the surface growth equation justified with the Eden model.
The origin of this term lies in non-equilibrium.  The third term is a
Langevin noise which mimics the stochastic nature of any growth
process and usually has a Gaussian distribution. In the last two
decades numerous studies came to light about the KPZ equation.
Without completeness we mention some of them. The basic physical
background of surface growth can be found in the book of Barab\'asi
and Stanley \cite{barab}. Later,  Hwa and Frey \cite{hwa1,hwa2}
investigated the KPZ model with the help of the  renormalization
group-theory and the self-coupling method which is a precise and
sophisticated method using Green's functions. Various dynamical
scaling forms of $C(x,t) = x^{-2\varphi} C(bx,b^zt)$  were
considered for the correlation function (where $\varphi, b $ and $z$
are real constants).  The field theoretical approach by L\"assig was
to derive and investigate the KPZ equation \cite{lass}.
Kriecherbauer and Krug wrote a review paper \cite{krug}, where the
KPZ equation was derived from hydrodynamical equations using a
general current density relation.

Several models exist and all lead to similar equations as the KPZ
model, one of them is the interface growth of bacterial colonies
\cite{Matsushita}. Additional general interface growing models were
developed based on the so-called Kuramoto-Sivashinsky (KS) equation
which shows similarity to the KPZ model with an extra $ \nabla^4 u$
term \cite{kuram1}, \cite{kuram2}.

Kersner and Vicsek investigated the traveling wave dynamics of the
singular interface equation \cite{kersner}  which is closely related
to the KPZ equation. One may find certain kind of analytic solutions
to the problem \cite{SaSp10} as already mentioned in \cite{CaDo11}.

\'Odor and co-worker intensively examined the two
dimensional KPZ equation with  dynamical simulations to investigate the aging properties of polymers or glasses \cite{odor}.

Beyond these continuous models based on partial differential
equations (PDEs), there are large number of purely numerical methods
available to study diverse surface growth effects. As a view we
mention the kinetic Monte Carlo \cite{mart} model, Lattice-Boltzmann
simulations \cite{sergi}, and the etching model \cite{melo}.

In this paper we investigate the solutions to the KPZ equation with
the traveling wave Ansatz in one-dimension applying various forms of
the noise term.  The effects of the parameters involved in the
problem are examined.

\section{Theory}
\label{sec:2}
In general, non-linear PDEs has no general mathematical theory which
could help us to understand general features or to derive physically
relevant solutions. Basically, there are two different trial
functions (or Ansatz) which have well-founded physical
interpretation. The first one is the traveling wave solution, which
mimics the wave property of the investigated phenomena described by
the non-linear PDE of the form
\begin{equation}
u(x,t) = f(x \pm c t)= f(\omega),
\label{self}
\end{equation}
where $c$ means the velocity of the corresponding wave. Gliding and
Kersner used the traveling wave Ansatz to investigate study numerous
reaction-diffusion equation systems \cite{glid}. To describe pattern
formation phenomena \cite{pattern} the traveling waves Ansatz is a
useful tool as well. Saarloos investigated the front propagation
into unstable states \cite{saar}, where traveling waves play a key
role.

This simple trial function can be generalized in numerous ways,
e.g., to $ e^{-\alpha t} f(x \pm ct):=e^{-\alpha t}  f(\omega) $
which describes exponential decay or to $  g(t) \cdot f(x \pm c\cdot
t):= g(t) f(\omega) $ which can even be a power law function of the
time as well. We note, that the application of these Ansatz to the
KPZ equation leads to the triviality of $ e^{-\alpha t} = g(t)
\equiv 1$.  In 2006, He and Wu developed the so-called exp-function
method \cite{he}  which relying on an Ansatz (a rational combination
of exponential functions),
involving many unknown parameters to be specified at the stage of solving the problem.
The method soon drew the attention of many researchers, who described it as “straightforward”, “reliable”, and
“effective”. Later, Aslan and Marinakis \cite{asl} summarized various applications of the Ansatz.

There is another existing remarkable Ansatz interpolating the
traveling-wave and the self-similar
Ansatz by Benhamidouche \cite{ben}.

The second one is the self-similar Ansatz \cite{sedov} of the form
$u(x,t)=t^{-\alpha}f\left(\frac{x}{t^\beta}\right):=t^{-\alpha}f(\omega). $
The associated mathematical and physical properties were
exhaustively discussed in our former publications \cite{barna}, \cite{barna2} or in a book chapter \cite{barna3} in the field of
hydrodynamics. All these kind of methods belong to the so-called
reduction mechanism, where applying a suitable variable
transformation the original PDEs or  systems of PDEs are reduced to
an ordinary differential equation (ODE) or  systems of ODEs.

\section{Results without the noise term}\label{sec:3}

Applying the traveling wave Ansatz to the KPZ PDE with $\eta(x,t) =
0 $, equation (\ref{kpz}) leads to the ODE of
\begin{equation}
-\nu f''(\omega) + f'(\omega)\left [ c-  \frac{\lambda}{2} f'(\omega) \right]   = 0,
\label{ODE_zajnelkul}
\end{equation}
From now on we use the Maple 12 mathematical program package to
obtain analytic solutions in closed forms. For equation
(\ref{ODE_zajnelkul}), it can be given as
\begin{equation}
f(\omega)   =  \frac{2}{\lambda} \ln \left(   \frac{\lambda
	\left[c_1  \nu e^{\frac{c\eta}{\nu}} +c_2 c  \right]}{2\nu c}
\right)\nu  , \label{zajnelkul_megoldas}
\end{equation}
where $c_1$ and $c_2$ are the constants of integration and $c$ is
the speed of the wave.

We fix this notation from now on throughout the paper. Note, that
this is an equation of a linear function $f(\omega) =  a\omega + b $
(just given in a complicated form) with any kind of parameter set,
except $c_1 = 0$ which gives a constant solution. This physically
means that there is a continuous surface growing till infinity which
is quite unphysical. Therefore, some additional noise is needed to
have surface growing phenomena. We remark  the general properties of
all the forthcoming solutions. Due to the Hopf-Cole transformation
\cite{hopf,cole} $(h = A \ln(y))$  convertes the non-linear KPZ
equation to the regular heat conduction (or diffusion) equation with
an additional stochastic source term eliminating the non-linear
gradient-squared term. All the solutions contain a logarithmic
function with a complicated argument. In this sense, the solutions
have the same structure, the only basic difference is the kind of
special function in the argument. If these argument functions take
periodically positive and negative real values then the logarithmic
function creates distinct intervals (small islands which describe
the surface growing mechanisms, and define the final solution). This
statement is generally true for our former study as well
\cite{imre}.

Remark that the solution to (\ref{kpz}) obtained from the
self-similar Ansatz reads
\begin{equation}
f(\omega) =  \frac{2\nu  } {\lambda} ln \left( \frac{\lambda c_1
	\sqrt{\pi\nu} \> \> erf[\omega/(2\sqrt{\nu})] + c_2}{2\nu}  \right),
\label{sol}
\end{equation}
where $erf[\ ]$ means the error function \cite{NIST}. Figure 1
compares these two solutions. We note the asymptotic convergence of
the self-similar solution and the divergence of the traveling-wave
solution. We have the same conclusion as in our former study
\cite{imre} (where the self-similar Ansatz was applied) , that
without any noise term the KPZ equation cannot be applied to
describe surface growth phenomena. The different kind of noise terms
define different kind of extra islands (parts of the solution having
compact supports) and these islands show a growth dynamics.

To have a better understanding between the two solutions, Fig. 2
shows the projection of both complete solutions
$u(x,y=0,t).$  The major differences are still present.


\begin{figure}
\begin{center}	
	\vspace*{-1.0cm}
	\includegraphics[keepaspectratio, width=9cm]{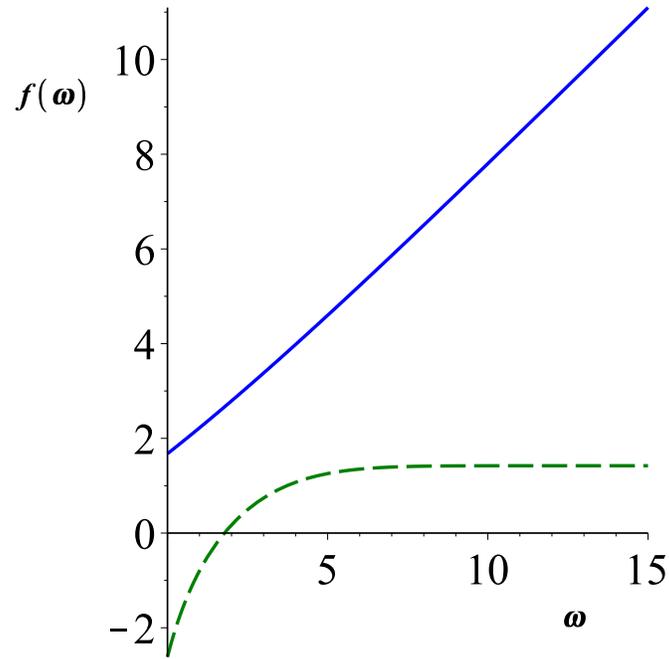}
	\vspace*{0.5cm} \caption{The two shape functions of the KPZ
		equation without any kind  of noise term. The solid line represents
		the solution for traveling-wave and the dashed line is for the
		self-similar Ansatz. The applied parameter set is $c_1 = c_2 = c =
		1,\ \nu =  4, \   \lambda = 3 $ }
	\label{ohne_zaj_1d}       
\end{center}
\end{figure}

\begin{figure}
	\begin{center}
	\includegraphics[keepaspectratio, width=9cm]{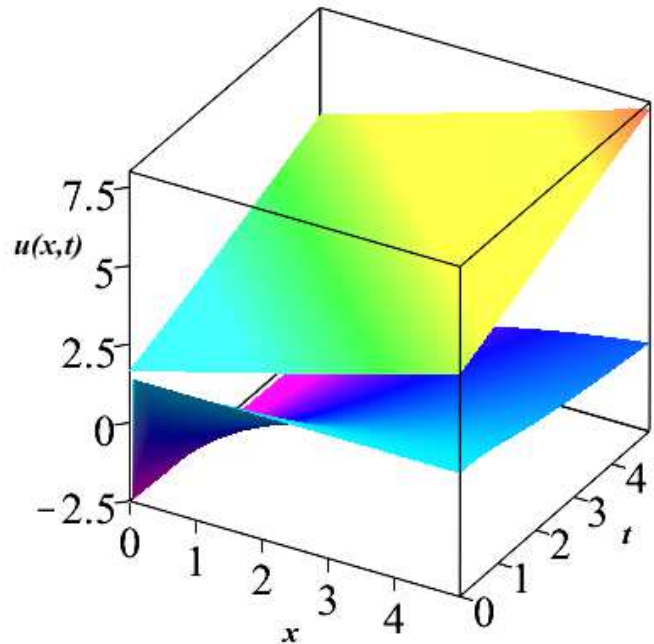}
	\vspace*{0.5cm}
	\caption{The two solutions of the KPZ equation without any noise term.
		The upper lying function represents the traveling-wave solution.
		The applied parameter set is the same as used above.}
	\label{ohne_zaj_3d}       
\end{center}
\end{figure}

\section{Results with various noise terms}\label{sec:4}

As we mentioned in our former study \cite{imre} only the additional
noise term makes the KPZ solutions interesting. We search the
solutions with the traveling-wave Ansatz, therefore is it necessary
that the noise  term $\eta$ should be an analytic function of $
\omega =x+ct$ like $ \eta(\omega) = a(x+ct)^2$. We will see that for
some kind of noise terms it is not possible to find a closed
analytic solution when all the physical parameters are free
$(\nu,\lambda,c,a)$, however, if some parameters are fixed it
becomes possible to find analytic expressions. It is also clear,
that it is impossible to perform a mathematically rigorous complete
function analysis according to all four physical and two integral
parameters $c_1, c_2$. We performed numerous parameter studies and
gave the most relevant parameter dependencies of the solutions.

\subsection{Brown noise $ n = -2$ }
As first, case let us consider the brown noise $\eta(x,t) =
\frac{a}{\omega^2 } $. It leads to the following ODE
\begin{equation}
-\nu f''(\omega) + f'(\omega)\left [ c-  \frac{\lambda}{2} f'(\omega) \right]  - \frac{a}{\omega^2} = 0.
\label{kpz_egy_per_etanegyzet}
\end{equation}

The solution can be given in the form
\begin{equation} f(\omega) =
\frac{1}{\lambda} \left( c\eta + \nu
\ln \left\{    \frac{ \lambda^2 \left [-c_1I_d \left( \frac{c\omega}{2\nu} \right)   + c_2K_d   \left(\frac{c\omega}{2\nu} \right) \right]^2}
{ c^2 \omega  \left[ K_d  \left(\frac{c\omega}{2\nu} \right) I_{d+1}
	\left(\frac{c\omega}{2\nu} \right) +   I_d
	\left(\frac{c\omega}{2\nu} \right) K_{d+1}
	\left(\frac{c\omega}{2\nu} \right)    \right]^2  } \right\} \right)
\end{equation} where $I_d(\omega)$ and  $K_d(\omega)$ are the modified Bessel
functions of the first and second kind \cite{NIST} with the
subscript of $d =  \frac{\sqrt{\nu^2-2a\lambda} }{2\nu} +1$. To
obtain real solutions for the KPZ equation (which provides the
height of the surface) the order of the Bessel function (notated as
the subscript) has to be non-negative and provides the following
constrain $\nu^2 \ge 2a\lambda $. This gives us a reasonable
relation among the three terms of the right hand side of equation
(\ref{kpz}). When the magnitude of the noise term $a$ becomes large
enough  no surface growth take place. Figure 3 presents  solutions
with different combinations of the integration constants $c_1, c_2$.
Having in mind, that the  $K_d()$ Bessel function of the second kind
is regular at infinity, one gets that it has a strong decay at large
argument $\omega$. The $c_1 = 0, c_2 = 0$ type solutions have
physical relevance. Figure 4 shows the complete solution of the KPZ
equation. It can be seen that a sharp and localized peak exists for
a short time. Therefore, no typical surface growth phenomena is
described with this kind of noise and initial conditions.

\begin{figure}
	\begin{center}
	\vspace*{1.0cm}
	\includegraphics[keepaspectratio, width=9cm]{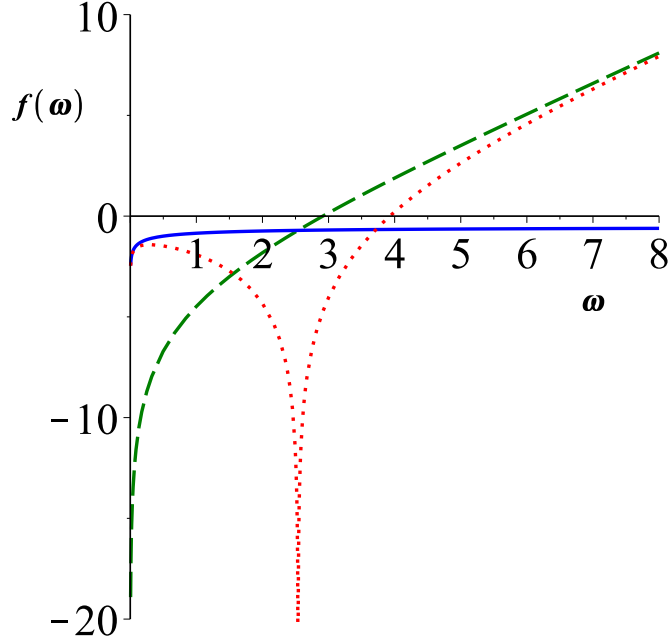}
	\vspace*{0.5cm} \caption{Three different shape functions for the
		brown noise $n=-2$. The applied physical parameter set is $ \lambda
		= 5, \nu = 3, a = 2$ and $ c = 2$. The dashed line is for $c_1 = 1,
		c_2 = 0$, the dotted line is for $ c_1 = c_2 =1$ and the solid line
		is for $c_1 = 0, c_2 = 1$, respectively.}
	\label{ohne_zaj_1d}       
\end{center}
\end{figure}
\begin{figure}
	\begin{center}
	\vspace*{0.1cm}
	\includegraphics[keepaspectratio, width=9cm]{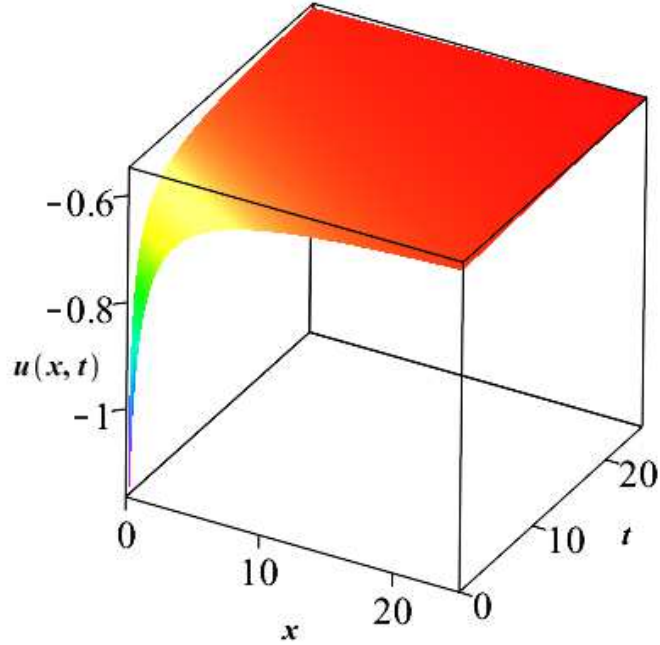}
	\vspace*{0.5cm} \caption{The solution $u(x,t)$ to the KPZ equation
		for the brown noise $n = -2$ with the parameter set of $c_1 = c_2 =
		c = 1, \ \nu =  4, \  \lambda = 3. $}
	\label{ohne_zaj_1d}       
\end{center}
\end{figure}


\subsection{Pink noise $ n = -1$ }
The noise term $\eta = \frac{a}{\omega}$  corresponds to the ODE
\begin{equation}
-\nu f''(\omega) + f'(\omega)\left [ c-  \frac{\lambda}{2} f'(\omega) \right]  -\frac{a}{\omega} = 0,
\label{kpz_egy_per_etanegyzet}
\end{equation}
whose general solution is
\begin{eqnarray}
f(\omega) = \frac{1}{\lambda} + 
\ln \left\{  \frac{-c\lambda [c_1M(\epsilon_{b}) -c_2U(\epsilon_{b})  ] }
{ M(\epsilon_{d})(2 \nu c U(\epsilon_{b})  +
	a \lambda U(\epsilon_{b})) +
	2M(\epsilon_{b})\nu c U(\epsilon_{d}) }
\right\},
\end{eqnarray}
where $M(\epsilon_{b})$ and $U(\epsilon_{d})$
are the Kummer M and Kummer U functions (for more see \cite{NIST})
with the parameters of $\epsilon_{b}=(\frac{2c\nu - a\lambda }{2c\nu},2,\frac{c\omega}{\nu})$ and $\epsilon_{d}=(\frac{ - a\lambda }{2c\nu},2,\frac{c\omega}{\nu})$. Figure 5 shows three different
shape functions corresponding to the pink noise. The evaluation of
direct parameter dependencies of the solutions are not trivial. In
some reasonable parameter range we found the following trends: for
fixed $a,c,\nu$ and larger $\lambda$ values, the solution shows more
independent well-defined "bumps" or islands and higher steepness of
the line which connects the maxima of the existing peaks of the
islands. At fixed  parameter values $a,c,\lambda$, different values
of $\nu$ just shift the position of the existing peaks. The role of
$a$ and $c$ is not defined. Figure 6 presents a total solution
$u(x,t)$ to the KPZ equation, the freely traveling three islands are
clearly seen.

\begin{figure}
	\begin{center}
		\includegraphics[keepaspectratio, width=9cm]{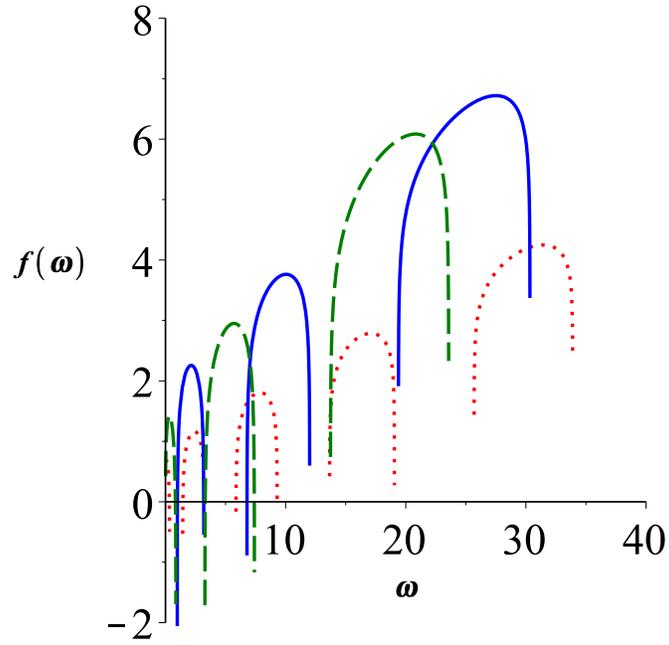}
		\vspace*{0.5cm}
		\caption{Three different shape functions for the pink noise  $(n = -1)$.
			Solid, dashed and dotted lines are for the parameter sets of
			$(c_1 = c_2 = 1; c = 1/2, \nu = 0.85, \lambda = 3,  a = 2)$,
			$(c_1 = c_2 = 1; c = 1/2, \nu = 0.85, \lambda = 2.5,  a = 2)$,
			$(c_1 = c_2 = 1; c = 0.6, \nu = 0.85, \lambda = 5,  a = 2)$,  respectively.
		}
	\end{center}
	\label{egy_per_omega}       
\end{figure}
\begin{figure}
	\begin{center}
		\includegraphics[keepaspectratio, width=9cm]{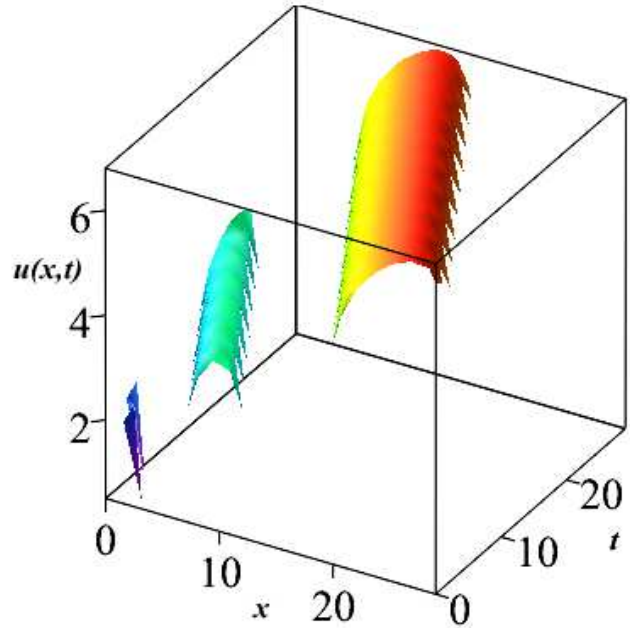}
		\vspace*{0.5cm}
		\caption{The total solution of the KPZ equation for $n = -1$  with
			the
			applied parameter set  $c_1 = c_2 = 1, c = 1/2,\nu =  0.85  \lambda = 3 $ and $a=2$. }
	\end{center}
	\label{egy_per_omega}       
\end{figure}


\subsection{White noise $ n = 0$ }
Here, the noise term is $\eta = a\omega^0 = a$ which leads to the
ODE of
\begin{equation}
-\nu f''(\omega) + f'(\omega)\left [ c-  \frac{\lambda}{2} f'(\omega) \right]  - a = 0,
\label{kpz_konstans}
\end{equation}

\begin{equation}
f(\omega) = \frac{\omega c}{\lambda} - \frac{\omega \sqrt{c^2-2a\lambda}}{\lambda} - \frac{2\nu \ln(2)}{\lambda} -
\frac{\nu \ln \left( \frac{c^2-2a\lambda}{  \lambda^2 \left [ c_1 e^{ \frac{\omega \sqrt{c^2-2a\lambda}}{\nu}}  -c_2 \right] ^2 }   \right) }{\lambda}
\end{equation}

Figure 7 shows two shape functions for two different parameter sets.
There exists basically two different functions depending on the
ratios of the integral constants $c_1$ and $c_2$. The first is a
pure linear function with infinite range and its domain represents
boundless surface growth, which is a physical nonsense. The second
solution is a sum of a linear and logarithmic function with a domain
bounded from above due to the argument of the $ln$ function. Figure
8 shows the final solution of the KPZ equation $u(x,t)$.  We note
that with the  substitution $\omega = x+ct$ only the first kind of
solution remains real. For the second parameter set which creates a
modified $ln$ function with a cut at well-defined argument becomes
complex.

\begin{figure}
	\begin{center}
	\includegraphics[keepaspectratio, width=9cm]{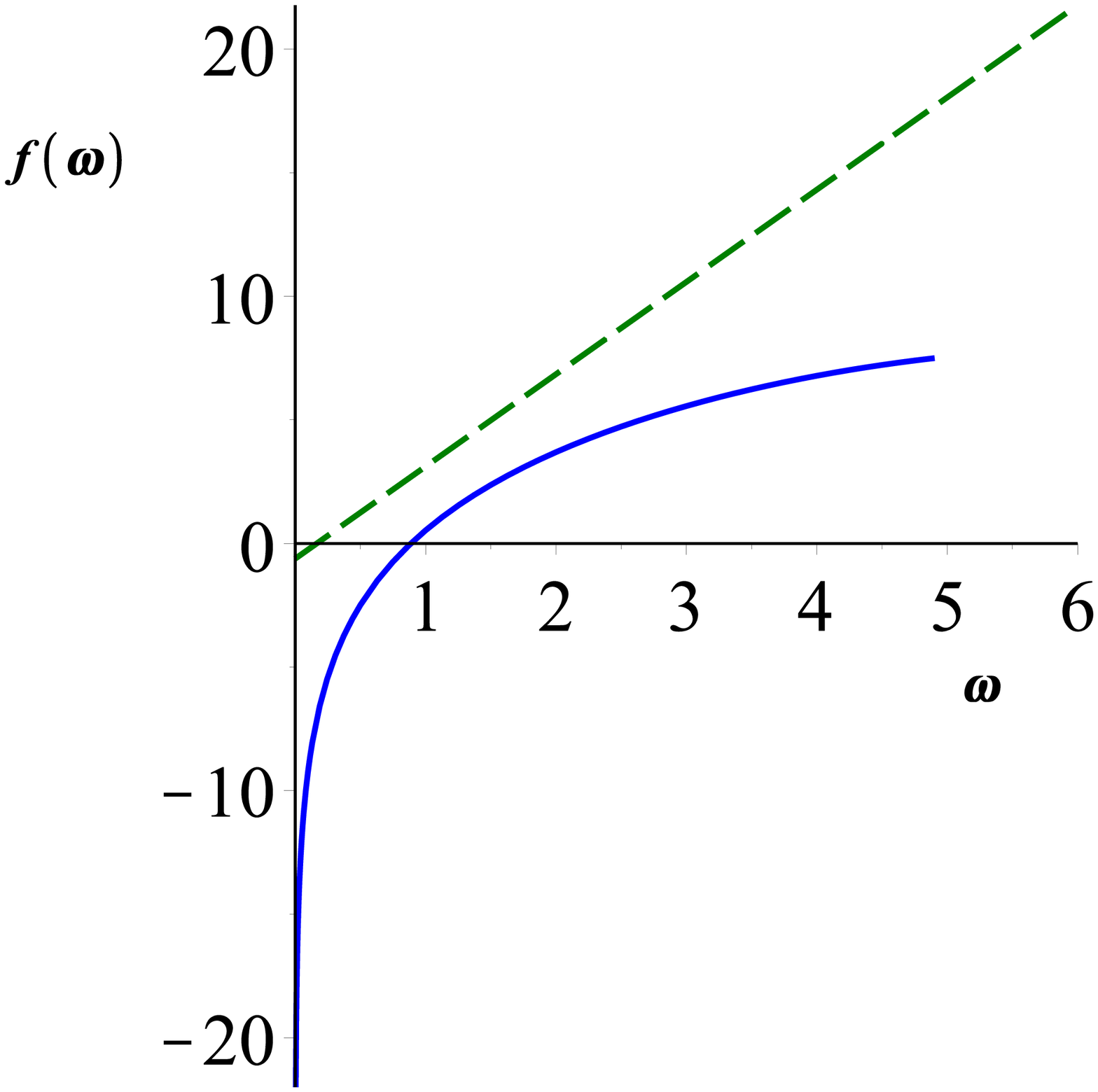}
	\vspace*{0.5cm}
	\caption{Two shape functions for the constant or white noise. The
		solid line is for the parameter set  $c_1 = 4,\ c_2 = -1,\ c = 0.3
		,\ \nu =  2,\  \lambda = 1,\ a = 1 $, and the dashed line is for
		$c_1 = c_2 = 1,\ c = 4,\ \nu =  0.5,\  \lambda = 1,\ a = 0.3 $,
		respectively.}
	\label{period_1d}       
\end{center}
\end{figure}
\begin{figure}
	\begin{center}
	\includegraphics[keepaspectratio, width=9cm]{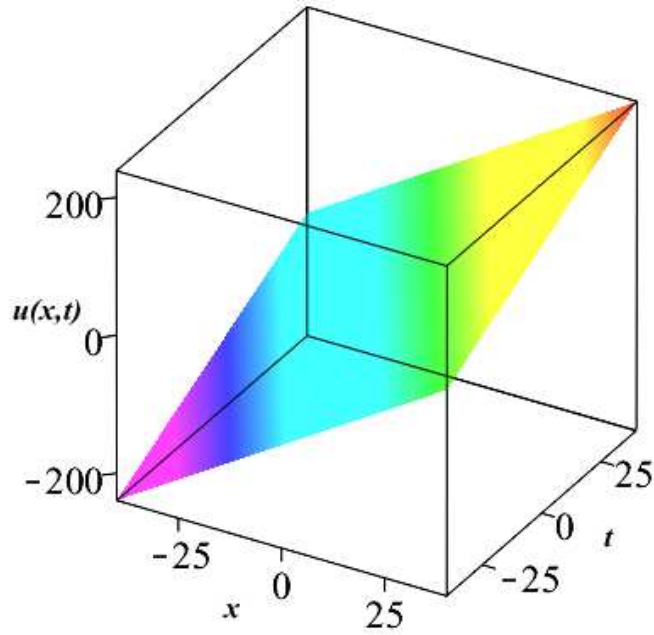}
	\vspace*{0.5cm}
	\caption{The KPZ solution for the constant or white noise. The
		applied parameter set is  $c_1 = c_2 = 1, c = 4,
		\nu =  0.5,  \lambda = 1, a = 0.3 $, respectively.}
	\label{period_1d}       
\end{center}
\end{figure}


\subsection{Blue noise $ n = 1$ }
The last color noise  $\eta = a\omega $ leads to the ODE of
\begin{equation}
-\nu f''(\omega) + f'(\omega)\left [ c-  \frac{\lambda}{2} f'(\omega) \right]  - a\omega  = 0,
\label{kpz_egy_per_etanegyzet}
\end{equation}
with the general solution of
\begin{equation}
f(\omega) =  \frac{c\omega}{\lambda} - \frac{4\nu \ln(2)}{3\lambda} +\frac{2\nu}{3\lambda} \ln \left\{ \frac{\lambda^2
	[ c_1Ai(\tilde{\omega})-c_2Bi(\tilde{\omega})]^3 }{\nu a [  Ai(1,\tilde{\omega})Bi(1,\tilde{\omega})  -Bi(1,\tilde{\omega})
	Ai(\tilde{\omega})) ]^3 } \right\} \label{13}
\end{equation}
where $Ai(\tilde{\omega}), Bi\tilde{\omega})$ denote the Airy
functions of the first and second kind and $Ai(1,\tilde{\omega})$
and $Bi(1,\tilde{\omega})$ are the first derivatives of the Airy
functions, where we used the following notation: $\tilde{\omega} =
\frac{-(2a\omega\lambda -c^2)4^{\frac{1}{3}} \left(
	\frac{a\lambda}{\nu^2} \right)^{1/3} }{4a\lambda}$. Exhaustive
details of the Airy function can be found in \cite{airy}. When
the argument $\omega$ is positive, $Ai(\omega)$ is positive, convex,
and decreasing exponentially to zero, while $Bi(\omega)$ is
positive, convex, and increasing exponentially. When $\omega$ is
negative, $Ai\omega)$ and $Bi\omega)$ oscillate around zero with
ever-increasing frequency and ever-decreasing amplitude.

Figure 9  represents shape functions with different parameter sets.
Our analysis showed that the composite argument of the $ln$ function
is purely real having a decaying oscillatory behavior with
alternatively positive and negative values.  The $ln$ function
creates infinite number of separate "bumps" or islands with compact
supports and infinite first spatial derivatives at their boarders.
Combining the first two terms of the (\ref{13}), we get an infinite
series of separate islands with increasing height. The ratio
$c/\lambda$ is the steepness of the line, this automatically defines
the steepness of the absolute height of the islands. The effects of
the various parameters are not quite independent and hard to define,
we may say that in general each parameter $\nu,\lambda,a,c $ alone
can change the widths, spacing and absolute height of the peaks.
Figure 10 shows the total solution of the KPZ equation. The
traveling "bumps" are clearly visible.

\begin{figure}
	\begin{center}
	\includegraphics[keepaspectratio, width=9cm]{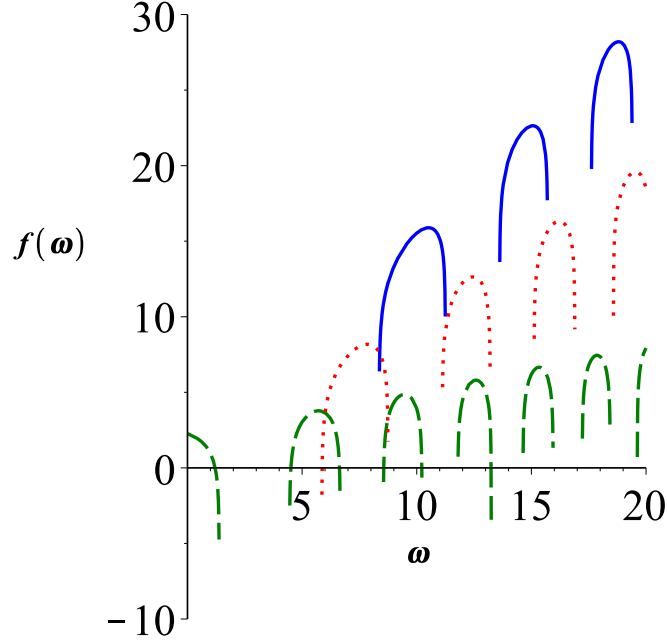}
	\vspace*{0.5cm}
	\caption{ The shape function for the blue noise for three parameter
		sets. The solid, dashed and dotted lines are for the parameter sets
		$(c_1 = 1, c_2 = 0, c = 3, a = 0.5, \nu = 1.5, \lambda = 2)$ ,
		$(c_1 = c_2 = c = 1, a = 1, \nu = 1, \lambda = 3)$ ,  and
		$(c_1 = c_2 = c = 1, a = 1, \nu = 1, \lambda = 3)$ ,  respectively. }
	\label{linear_1d_fig}       
\end{center}
\end{figure}
\begin{figure}
	\begin{center}
		\includegraphics[keepaspectratio, width=9cm]{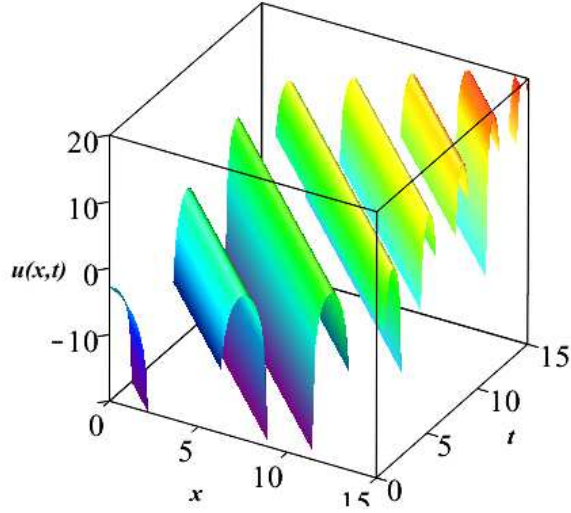}
		\vspace*{0.5cm} \caption{The   solution $u(x,t)$ for the $n = 1$ or
			blue noise  with the
			applied parameter set of $c_1 = c_2 = c = 1,\nu =  2,  \lambda = 3, a = 1 $}
		\label{linear_3d}       
	\end{center}
\end{figure}


\subsection{Lorentzian noise }
As a first non-colour noise let us consider the Lorentzian noise of
the form  $\eta = \frac{a}{1+\omega^2}$. It leads to the ODE of
\begin{equation}
-\nu f''(\omega) + f'(\omega)\left [ c-  \frac{\lambda}{2} f'(\omega) \right]  - \frac{a}{1+\omega^2}   = 0,
\label{kpz_egy_per_etanegyzet}
\end{equation}

We mention, that for the classical exponential and Gaussian noise
distributions we could not give solutions in closed analytic form.
Unfortunately, there is no closed analytic expression available if
all the parameters $(\nu,\lambda,c,a)$ are free. The formal solution
contains integrals of the Heun  C confluent functions multiplied by
some polynomials. However, if the parameters $a,\lambda,\nu $ are
fixed, there is analytic solution available  for free propagation
speed $c$.
The exact solution for $a = \lambda = \nu = 1/2,$ and $ c = 2$ is
the following
\begin{eqnarray}
f(\omega) = c \omega + \nonumber \\ 
2 \ln \left\{  \frac{ c_1C( B ) -c_2 \omega	C(A) }
{  2(\omega^4 + \omega^2)[C (A) C'(B) -  C (B) C'(A)] + (1+\omega^2) C (A) C (B)} \right\},
\end{eqnarray}

where $C'()$ means the first derivative of the Heun  C function. For
the better transparency we introduce the following notations $
A=0,\frac{1}{2},1,\frac{c^2}{4},1-\frac{c^2}{4};-\omega^2$ and
$B=0,-\frac{1}{2},1,\frac{c^2}{4},1-\frac{c^2}{4};-\omega^2$.

Figure 11 shows the shape function for given parameter set. There is
a broad island close to the origin and numerous tiny ones at larger
arguments.  The numerical accuracy of Maple 12 was enhanced to reach
this resolution.  It is well-known that the Heun functions are the
most complicated objects among special functions and the evaluations
needs more computer time.

Figure 12 presents the total solution of the original KPZ. Due to
the substitution  $ \omega = x + ct$ the original local solution
broke down to several smaller islands which freely propagate in time
and space.
\begin{figure}
	\begin{center}
		\hspace*{0.4cm}
		\includegraphics[keepaspectratio, width=9cm]{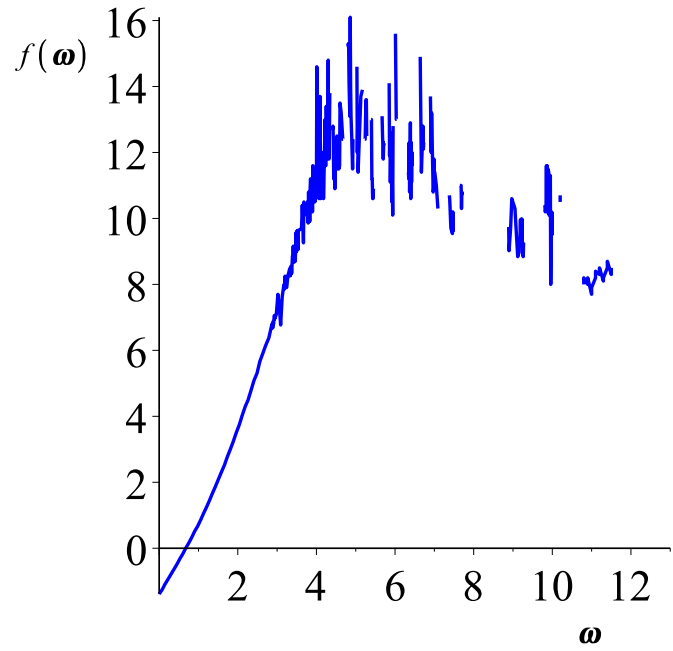}
		\vspace*{0.5cm} \caption{The shape function for the Lorentzian
			noise.  The applied parameters are  $c_1  =0.5,  c_2 = 2, c = 1, \nu
			= 1, a = 1,  \lambda =  3,$}
		\label{period_1d}       
	\end{center}
\end{figure}
\begin{figure}
	\begin{center}
		\hspace*{0.4cm}
		\includegraphics[keepaspectratio, width=9cm]{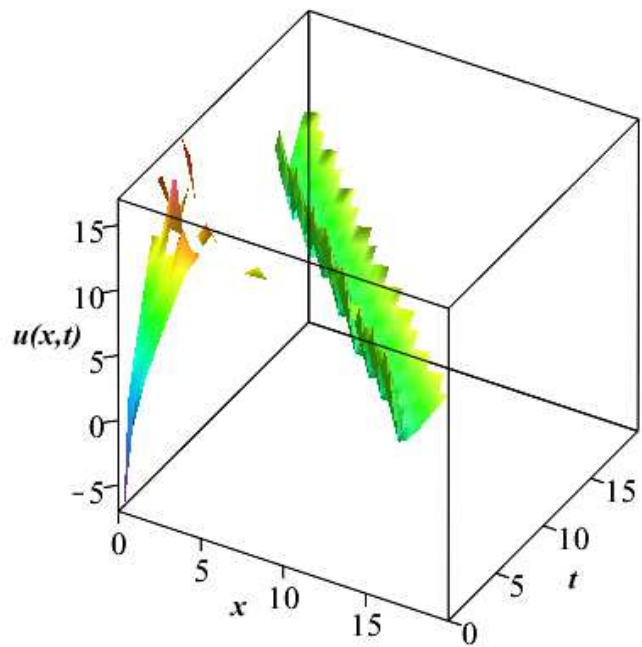}
		\vspace*{0.5cm} \caption{The solution of the KPZ equation for
			Lorentzian noise,  with the parameters mentioned above.}
		\label{period_3d}       
	\end{center}
\end{figure}


\subsection{Periodic noise }
The last   perturbation investigated is a periodic function $\eta =
a \sin(\omega)$ and
\begin{equation}
-\nu f''(\omega) + f'(\omega)\left [ c-  \frac{\lambda}{2} f'(\omega) \right]  - a\sin(\omega)  = 0.
\label{kpz_egy_per_etanegyzet}
\end{equation}
The general solution can be given as
\begin{equation}
f (\omega) =  \frac{1}{\lambda}  \left ( c\omega    + 2 \ln \left\{
\frac{\lambda[c_1C(\epsilon_{a}) - c_2 S(\epsilon_{a}
	)]} { \nu[-C'(\epsilon_{a}) S(\epsilon_{a}) +
	C(\epsilon_{a}) S'(\epsilon_{a})]}   \right\}  \right),
\label{17}
\end{equation}
where $C(\epsilon_{a}),S(\epsilon_{a}),
C'(\epsilon_{a}) $ and $S'(\epsilon_{a})$ are the
Mathieu S and Mathieu C functions and the first derivatives. For
basic properties we refer to \cite{NIST}. For a complex study about
Mathieu functions see \cite{mat1,mat2,mat3}. In (\ref{17}), we used
the abbreviation of $\epsilon_{a} = - \frac{c^2}{\nu^2},-\frac{a\lambda}{\nu^2},-\frac{\pi}{4} + \frac{\omega}{2}$.

Figure 13 shows a typical shape function for the periodic noise
term. Due to the elaborate properties of even the single Mathieu C
or S functions for some parameter pairs $a,q$ the function is finite
with periodic oscillations and for some neighboring parameters it is
divergent for large arguments. No general parameter dependence can
be stated. The parameter space of the set of six real values
$(c_1,c_2,c,a,\nu,\lambda)$ is too large to map. After the
evaluation of numerous shape functions we may state, that a typical
shape function is presented with two larger islands close to the
origin and numerous smaller intervals. For large argument $\omega$
the shape function shows a steep decay.

Figure 14 shows the complete solution. Note, that the first two
broader islands can be seen as they freely travel. Due to the finite
resolution the smaller islands are represented as irregular noise in
the background.

\begin{figure}
	\begin{center}
		\hspace*{0.4cm}
		\includegraphics[keepaspectratio, width=9cm]{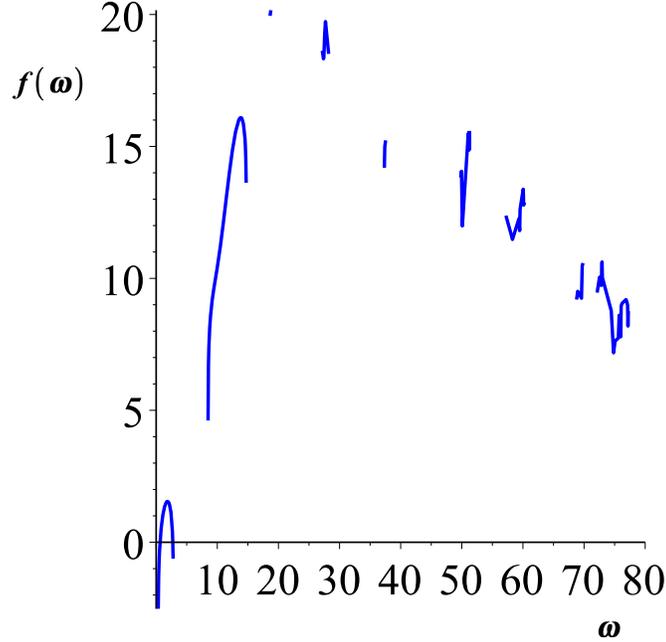}
		\vspace*{0.5cm}
		\caption{The shape function for the periodic noise. The applied parameters are  $c_1  =0.5,  c_2 = 2, c = 1, \nu = 1, a = 1,  \lambda =  3,$}
		\label{period_1d}       
	\end{center}
\end{figure}

\begin{figure}
	\begin{center}
		\vspace*{-1.4cm}
		\includegraphics[keepaspectratio, width=9cm]{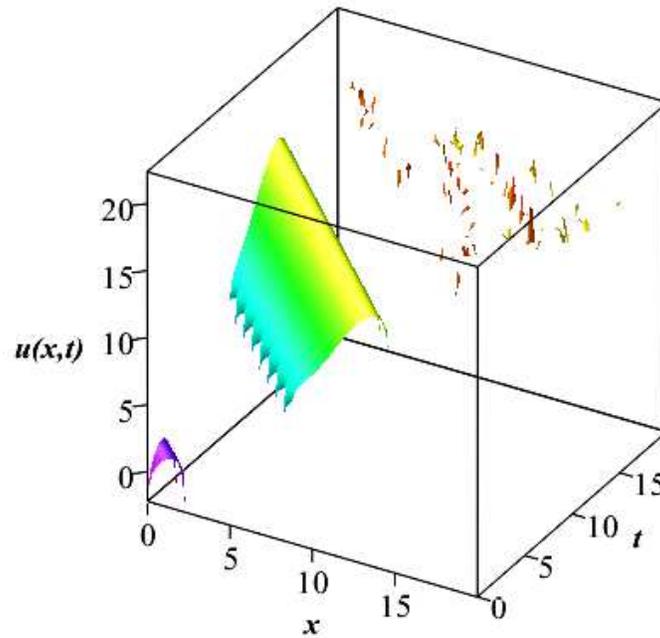}
		\vspace*{0.5cm}
		\caption{The complete traveling wave solution $ u(x,t) $ for periodic noise with the
			same parameter set as given above.}
		\label{period_3d}       
	\end{center}
\end{figure}

\section{Conclusions}
In summary, we can say that with an appropriate change of variables
applying the self-similar Ansatz one may obtain analytic solution
for the KPZ equation for one spatial dimension with numerous noise
terms. We investigated four type of power-law  noise $\omega^n$ with
exponents of $-2,-1,0,1$, called the brown, pink, white and blue
noise, respectively. Each integer exponent describes completely
different dynamics. Additionally, the properties of Gaussian and
Lorentzian noises are investigated. Providing completely dissimilar
surfaces with growth dynamics. All solutions can be described with
non-trivial combinations of various special functions, like error,
Whittaker, Kummer or Heun. The parameter dependencies of the
solutions are investigated and discussed. Future works are planned
for the investigations of  two dimensional surfaces.

\section*{Acknowledgment}

This work was supported by project no. 129257 implemented with the support provided from the National Research, Development and Innovation Fund of Hungary, financed under the $K \_ 18$ funding scheme.
The described study was carried out as part of the EFOP-3.6.1-16-00011 “Younger and Renewing University – Innovative Knowledge City – institutional development of the University of Miskolc aiming at intelligent specialization” project implemented in the framework of the Szechenyi 2020 program. The realization of this project is supported by the European Union, co-financed by the European Social Fund.

\bibliographystyle{unsrt}  
\bibliography{references}  






\end{document}